\title{SPIN STIFFNESS \\
IN \\ THE HUBBARD MODEL}
\author{J.M.J. van Leeuwen, M.S.L. du Croo de Jongh, \\
        and P.J.H. Denteneer\\
        Instituut-Lorentz, University of Leiden,\\
        P. O. Box 9506, 2300 RA Leiden, The Netherlands}
\date{}
\documentclass[12pt]{article}
\usepackage{bbm}
\newcommand{\be}{\begin{equation}}
\newcommand{\ee}{\end{equation}}
\newcommand{\bc}{\begin{center}}
\newcommand{\ec}{\end{center}}
\newcommand{\ha}{{\scriptstyle \frac{1}{2}}}

\newcommand{\fj}{\varphi_j}

\newcommand{\bk}{{\bf k}}

\newcommand{\bq}{{\bf q}}
\newcommand{\br}{{\bf r}}
\newcommand{\bs}{{\bf s}}
\newcommand{\bA}{{\bf A}}
\newcommand{\bJ}{{\bf J}}
\newcommand{\bM}{{\bf M}}
\newcommand{\bN}{{\bf N}}
\newcommand{\bQ}{{\bf Q}}
\newcommand{\bsig}{\vec{\sigma}}
\newcommand{\thz}{\vec{\vartheta}^z}
\newcommand{\thx}{\vec{\vartheta}^x}
\newcommand{\dthz}{\dot{\vec{\vartheta^z}}}
\newcommand{\Hh}{{\cal H}_h}
\newcommand{\Htwo}{\overline{\overline{\cal H}}_2}
\newcommand{\Hone}{\vec{\cal H}_1}
\newcommand{\Jone}{\overline{\overline{\bJ^1}}}
\newcommand{\Jhz}{\bJ^z_h}
\newcommand{\Ms}{M_{\rm st}}
\newcommand{\ms}{m_{\rm st}}
\newcommand{\rs}{\rho_{\rm s}}
\newcommand{\rsbb}{\overline{\overline{\rho}}_s}
\newcommand{\chbb}{\overline{\overline{\chi}}}
\newcommand{\Ibb}{\overline{\overline{I}}}
\newcommand{\tr}{{\rm tr}\,}

\begin{document}
\hoffset=-0.0 true cm
\voffset=-2.0 true cm
\maketitle
\vspace{1cm}
\begin{abstract}
The spin stiffness $\rs$ of the repulsive Hubbard model that occurs
in the hydrodynamic theory of antiferromagnetic spin waves is shown
to be the same as the thermodynamically defined stiffness involved
in twisting the order parameter. New expressions for $\rs$ are derived,
which enable easier interpretation, and connections with superconducting
weight and gauge invariance are discussed.
\end{abstract}
\vspace{1.0cm}
Short title: Spin stiffness in the Hubbard model
\newpage

\section{Introduction}
\renewcommand{\theequation}{1.\arabic{equation}}
\setcounter{equation}{0}
When a continuous symmetry is broken a Goldstone mode appears,
which has to be considered as a slow variable and which necessitates
an extension
of the hydrodynamic equations \cite{Fors}.
The most famous example of this phenomenon is the case of superfluidity,
where the Goldstone mode leads to second sound.
The second sound velocity is governed by the superfluid density
$\rs$, which is a measure of the stiffness of the superfluid
order parameter against spatial variations. This situation has
a perfect analogy in antiferromagnetic ordering, where the
Goldstone mode is a spin wave and $\rs$ a spin stiffness.
A third member in this family is superconductivity, where $\rs$ may be
viewed as density of superconducting carriers. In the last example,
one can relate $\rs$ also to the response to an electromagnetic field:
it is inversely proportional to the square of the
London penetration depth in the
Meissner effect \cite{SWZ}.

In this paper, we analyze the antiferromagnetic ordering and the
associated spin waves in the context of the Hubbard model. However,
our considerations are applicable to a larger class of quantum
lattice models; the Hubbard model merely serves as an example and
an opportunity to give explicit formulae. The purpose of reconsidering
this well-known theory is that it enables us to establish, on
a microscopic basis, the general connection between spin-wave
velocity and the stiffness $\rs$ \cite{HalHo}.
 In doing so, we find new expressions
for $\rs$ which reveal more clearly the nature of $\rs$ than does the
rather formal definition in terms of a twist in the order parameter
\cite{FBJ}. By using the Hubbard model as an example we can easily
make the connection with the superconducting language, by means
of a transformation between repulsive and attractive Hubbard models.
This gives us the opportunity to clarify some misunderstanding in
the literature concerning gauge invariance.
In general, the spirit of the paper is an articulation
for the Hubbard model of the
analysis by Forster \cite{Fors} of the connection between broken symmetry,
correlation functions and hydrodynamics.

We have organised the paper as follows: we start with a brief discussion
of the tools needed for the analysis. Then we show how $\rs$ enters
in the (hydrodynamic) equations for the spin waves. In section 4,
we work out the (thermodynamic) definition of $\rs$ in terms of
a twist in the order parameter and subsequently we demonstrate
in section 5 that the hydrodynamic $\rs$ is indeed the spin stiffness.
In section 6, we draw the analogy with $\rs$ as the response function
between the current and the inducing gauge field as it appears in
the theory of superconductivity. In section 7, we comment on formal
manipulations with the expression for $\rs$ and derive new expressions
which are more transparant. We close with a brief discussion.

\section{Preliminaries}
\renewcommand{\theequation}{2.\arabic{equation}}
\setcounter{equation}{0}

\subsection{The Hubbard model}
The Hubbard model is represented by the hamiltonian \cite{Hub}: \\
\begin{equation}
\Hh = - t \sum_{j,\delta,\sigma } c_{j+\delta,\sigma}^{\dagger}
c_{j \sigma}
+ U \sum_{j}n_{j \uparrow}n_{j \downarrow} ,   \label{eq:Hh}
\end{equation}
where $c_{j \sigma}^{(\dagger)}$ are electron
annihilation (creation) operators for
site $j$ with spin $\sigma$ ($\sigma = \uparrow, \downarrow$ corresponds
to $+1, -1$, respectively, if it does not occur as an index). Neighboring
sites of site $j$ are denoted by $j+\delta$. We consider a bipartite lattice,
such that hops from site $j$ to site $j+\delta$ are always from one
sublattice to the other. $t$ is the exchange- or hopping integral,
$U$ is the on-site interaction between electrons of opposite spin, and
$n_{j \sigma}=c_{j \sigma}^{\dagger}c_{j \sigma}$ is the occupation
number operator.

$\Hh$ has a rich set of symmetries. For our purpose it
suffices to mention the magnetic symmetry:
\be
\left[ \Hh , \bM \right] = 0 ~,
\ee
where $\bM$ is the total magnetization operator:
\be
\bM = \sum_{j} \bs_j ~,
\ee
with $\bs_j$ the local magnetization operator:
\be
\bs_j = \ha \sum_{\sigma,\sigma'}
c_{j \sigma}^{\dagger}\bsig_{\sigma\sigma'}c_{j \sigma'} ~,
\ee
with $\bsig$ the set of the three Pauli matrices.

\subsection{Averages and operator inner products}
Averages are based on a (grand) canonical ensemble described by a
hamiltonian ${\cal H}$. The partition function gives the free energy
$F$ according to:
\be
F = -\frac{1}{\beta} \ln \tr {\rm e}^{-\beta {\cal H}} ~,
\ee
where $\beta = 1/k_{\rm B}T$. Averages are taken with respect to the
canonical weight:
\be
\langle {\cal A} \rangle =
{\rm e}^{\beta F}\, \tr {\rm e}^{-\beta {\cal H}} {\cal A} ~. \label{eq:ave}
\ee
We will need an inner product in operator space, for which we take
\cite{Kubo,Fors}:
\be
 \left( {\cal A},{\cal B} \right) =
\frac{1}{\beta} \int_0^{\beta} {\rm d} \lambda
\left\{ \langle {\cal A}^{\dagger} {\rm e}^{-\lambda {\cal H}} {\cal B} \,
{\rm e}^{\lambda {\cal H}} \rangle - \langle {\cal A}^{\dagger} \rangle
\langle {\cal B} \rangle \right\} ~. \label{eq:inp}
\ee
In addition to having all the properties of an inner product this definition
has a few more desirable aspects. If the hamiltonian $\cal H$ is perturbed
by adding the operator $\delta {\cal H}$, the linear response of an operator
$\cal J$ is given by:
\be
 \delta \langle {\cal J} \rangle  \equiv
\langle {\cal J} \rangle_{{\cal H} + \delta {\cal H}} -
\langle {\cal J} \rangle =
- \beta \left( \delta {\cal H}^{\dagger}, {\cal J} \right) ~.
\label{eq:delJ}
\ee
The change in the free energy is to second order in $\delta {\cal H}$:
\be
 \delta F  = \langle \delta {\cal H} \rangle - \frac{\beta}{2}
\left(\delta {\cal H}^{\dagger}, \delta {\cal H} \right) . \label{eq:delF}
\ee
The last two equations are derived in the Appendix.

Another property of the inner product, that we will use, concerns
the Heisenberg operators:
\be
{\cal A}(t) = {\rm e}^{it{\cal H}/\hbar} {\cal A} \,{\rm e}^{-it{\cal H}/\hbar}
\ee
and their time derivatives
\be
\dot{\cal A}(t) = \frac{i}{\hbar} \left[ {\cal H} , {\cal A}(t) \right] ~.
\ee
Using the invariance of the trace under cyclic permutation of the operators
one easily proves the relation:
\be
\left( {\cal B}, \dot{\cal A}(t) \right) = \frac{i}{\beta\hbar}
\langle \left[ {\cal B}^{\dagger}, {\cal A}(t) \right] \rangle ~.
\label{eq:timederid}
\ee

\subsection{Hydrodynamic equations}
Linear hydrodynamic equations are obtained by first identifying
the set of hydrodynamic variables ${\cal A}_j$ \cite{Fors}.
They span a hydrodynamic subspace (using the operator inner product
defined above). Projection on the hydrodynamic subspace is achieved
by the projection operators:
\be
 \mathbbm{P} {\cal O} = \sum_{j,m} \, {\cal A}_j P_{jm}
\left( {\cal A}_m , {\cal O} \right)  ~,
\ee
where the matrix $P$ is the inverse of the matrix of susceptibilities
$\left( {\cal A}_i , {\cal A}_j \right)$:
\be
\sum_m \, P_{jm} \left( {\cal A}_m , {\cal A}_k \right) = \delta_{jk} ~,
\ee
where all indices run through the hydrodynamic subspace.

The dissipation-free hydrodynamic equations result from the
projections on the hydrodynamic subspace of the time derivatives:
\be
 \mathbbm{P} \dot{\cal A}_\ell = \sum_j \, {\cal A}_j \, \Omega_{j\ell} ~,
\ee
with $\Omega_{j\ell}$\, given by
\be
 \Omega_{j\ell} = \sum_m \, P_{jm}
\left( {\cal A}_m , \dot{\cal A}_\ell \right) \label{eq:Omjl}
\ee
The hydrodynamic equations then read:
\be
 \frac{\rm d}{{\rm d}t} \langle {\cal A}_\ell (t) \rangle =
\sum_j \, \langle {\cal A}_j (t) \rangle \, \Omega_{j\ell} ~.
\ee

\section{Spin stiffness and hydrodynamic spin waves}
\renewcommand{\theequation}{3.\arabic{equation}}
\setcounter{equation}{0}
The slow or hydrodynamic modes derive from the conserved quantities
and the Goldstone modes of a broken symmetry \cite{Fors}.
Since the total magnetization is conserved, the fourier components
\be
 \bM  (\bk) = \sum_j \, {\rm e}^{i\bk\cdot\br_j} \bs_j
\ee
are slow modes for small $\bk$. The symmetry breaking that we consider is
antiferromagnetic ordering, which occurs in the Hubbard model at
low temperatures near half-filling.
The staggered magnetization
\be
 \bN = \sum_j \, (-1)^j \, \bs_j
\ee
then acquires a non-zero average. $(-1)^j$ equals $+1$ on one
sublattice and $-1$ on the other sublattice. $\bN$ can be seen
as a fourier component $\bM (\bQ)$, where the wave vector $\bQ$
has the property:
\be
{\rm e}^{i\bQ\cdot\br_j} = (-1)^j
\ee
The fourier components $\bN (\bk) = \bM (\bk + \bQ)$ are the slow modes
for small $\bk$ associated with the antiferromagnetic symmetry breaking.
We assume that the system orders antiferromagnetically in the
$x$-direction, i.e.
\be
\langle N^{x} \rangle = M_{\rm st} \neq 0 ~.
\ee
The spin waves arise from an interplay of the small-$\bk$ components
of $\bM (\bk)$ and $\bN (\bk)$. Which components play a role can be seen
from the general commutation relation
\be
\left[ M^{\alpha} (\bk) , M^{\beta} (\bk') \right] =
i\, M^{\gamma} (\bk + \bk') ~, \label{eq:comM}
\ee
with $\alpha, \beta, \gamma$ a cyclic permutation of $x, y, z$.
Taking $\gamma =x$, we see that a non-zero average of $N^x$ couples
$M^z(\bk)$ to $N^y(\bk)$ and $M^y(\bk)$ to $N^z(\bk)$. The two pairs
are equivalent and we focus on the first couple. In principle, one should
write down the hydrodynamic equations in the combined 6-dimensional
space of $\bM (\bk)$ and $\bN (\bk)$, but symmetry considerations
permit to restrict the equations to the $2 \times 2$ subspace
of $M^z(\bk)$ and $N^y(\bk)$.
So we obtain equations of the form \cite{Fors,Kruger}:
\begin{eqnarray}
\frac{\partial \langle M^z(\bk,t) \rangle}{\partial t} & = &
\langle M^z(\bk,t) \rangle \, \Omega_{zz} +
\langle N^y(\bk,t) \rangle \, \Omega_{yz} ~,  \label{eq:hydeq1} \\
\frac{\partial \langle N^y(\bk,t) \rangle}{\partial t} & = &
\langle M^z(\bk,t) \rangle \, \Omega_{zy} +
\langle N^y(\bk,t) \rangle \, \Omega_{yy}  \label{eq:hydeq2}
\end{eqnarray}
The $\Omega_{j\ell}$ are calculated with (\ref{eq:Omjl}). The entry
$\left( {\cal A}_m , \dot{\cal A}_\ell \right)$ is simple, as it can
be calculated with (\ref{eq:timederid}) and (\ref{eq:comM}):
\begin{eqnarray}
 \left( M^z(\bk), \dot{M}^z(\bk) \right) & = & 0 ~~~~~~~~~~~~~
 \left( M^z(\bk), \dot{N}^y(\bk) \right)  =  \Ms/\beta\hbar \\
 \left( N^y(\bk), \dot{M}^z(\bk) \right) & = & -\Ms/\beta\hbar ~~~
 \left( N^y(\bk), \dot{N}^y(\bk) \right)  =  0 ~. \label{eq:NyMzd}
\end{eqnarray}
The matrix elements $P_{jm}$ are obtained from the inverse of
$\left( {\cal A}_m , {\cal A}_k \right)$, which in turn is interpreted
as the response to a perturbation as in (\ref{eq:delJ}). So we consider
first a perturbation:
\be
 \delta {\cal H} = - \xi \, M^z(\bk) ~, \label{eq:pertMz}
\ee
with a small amplitude $\xi$. It will induce:
\be
 \delta \langle M^z(\bk) \rangle =
\beta\xi \left( M^z(\bk), M^z(\bk) \right) = N \chi_{\perp}(\bk) \xi ~,
\ee
where $N$ is the number of sites.
In the second equality we introduce a susceptibility $\chi_\perp$, where
the subscript is a reminder that it concerns a susceptibility
perpendicular to the order parameter direction. We may drop the
$\bk$-dependence since we are interested in small $\bk$ and
$\chi_\perp(\bk)$ is expected to be regular
for small $\bk$.

For symmetry reasons the perturbation (\ref{eq:pertMz}) will give no
effect $\delta \langle N^y(\bk) \rangle$. So,
\be
 \left( M^z(\bk), N^y(\bk) \right) =
\left( N^y(\bk), M^z(\bk) \right)^\ast = 0   ~.
\ee
The perturbation,
\be
 \delta {\cal H} = - \xi \, N^y(\bk) ~, \label{eq:pertNy}
\ee
requires a different treatment. It tends to turn over the spontaneous
order from the $x$-direction to the $y$-direction. For small $\bk$
the response will diverge and we will determine the singular
behavior in section 5. For the moment, we use the result,
\be
 \left( N^y(\bk), N^y(\bk) \right) = \frac{\Ms^2}{\beta N\rs k^2}
 \left( 1 + {\cal O}(k) \right)  ~~~~\bk \rightarrow 0 ~. \label{eq:divk2}
\ee
One can see (\ref{eq:divk2}) as the {\em hydrodynamic} definition of the
spin stiffness $\rs$.

Now all the ingredients for $\Omega_{j\ell}$ are present and we find:
\begin{eqnarray}
 \Omega_{zz} & = & 0 ~~~~~~~~~~~~~~~
\Omega_{zy} = \frac{\Ms}{\hbar N \chi_\perp}   \\
 \Omega_{yz} & = & -\frac{N\rs k^2}{\Ms \hbar} ~~~~~ \Omega_{yy} = 0
\end{eqnarray}
The hydrodynamic equations (\ref{eq:hydeq1})-(\ref{eq:hydeq2})
have a solution:
\be
\langle M^z(\bk,t) \rangle = \langle M^z(\bk,0) \rangle
{\rm e}^{\pm ickt}   ~~~~~~~
\langle N^y(\bk,t) \rangle = \langle N^y(\bk,0) \rangle
{\rm e}^{\pm ickt}  ~,
\ee
with the spin wave velocity $c$ given by:
\be
 c = \frac{1}{\hbar} \sqrt{\frac{\rs}{\chi_\perp}} ~, \label{eq:HHrel}
\ee
a relation due to Halperin and Hohenberg \cite{HalHo}.

\section{Spin stiffness and twisted order parameter}
\renewcommand{\theequation}{4.\arabic{equation}}
\setcounter{equation}{0}
The more fundamental definition of the spin stiffness $\rs$ relates it
to the increase in the free energy due to imposing a twist in the
boundary condition on the order parameter \cite{FBJ}.
E.g., one can have the order parameter point in the $x$-direction
at one end of the system and in the $y$-direction at the other end.
For a continuous symmetry the twist in the order parameter will spread
itself equally over the system. So we imagine that the order parameter
slowly rotates around the $z$-axis:
\be
\langle s^x_j \rangle =
\ms \cos\left(\left( \bq + \bQ \right) \cdot \br_j \right) ~~~~~~~~~~~~
\langle s^y_j \rangle =
\ms \sin\left(\left( \bq + \bQ \right) \cdot \br_j \right) ~, \label{eq:sxsy}
\ee
where $\bq$ is the pitch of the twist and $\ms = \Ms/N$. The two equations
(\ref{eq:sxsy}) can be combined into:
\be
\langle s^+_j \rangle =
\ms \exp\left( i\left( \bq + \bQ \right) \cdot \br_j \right) ~,
\label{eq:s+tw}
\ee
with
\be
 s^+_j = s^x_j + i s^y_j = c_{j \uparrow}^{\dagger}c_{j \downarrow} ~,
\ee
the spin-raising operator. The condition (\ref{eq:s+tw}) leads to an
increment in the free energy which for small $\bq$ can be written as:
\be
 F(\bq) = F(0) + \ha N \rs q^2 + \cdots ~. \label{eq:Fq}
\ee
The term linear in $\bq$ is absent due to inversion symmetry.
Formula (\ref{eq:Fq}) has to be considered as the {\em thermodynamic}
definition of the spin stiffness $\rs$ against a twist.

This formal definition does not lead to a simple calculational
scheme since it involves the computation of the free energy
of a spatially inhomogeneous system. We can however map the system
with the condition (\ref{eq:s+tw}) on a different system with
an easier constraint by an unitary transformation ${\cal U}$ of
the form:
\be
 {\cal U} = {\rm e}^{i\bq\cdot\sum_j \, \br_j s^z_j}  ~. \label{eq:U}
\ee
${\cal U}$ transforms ${\cal H}$ and the density matrix $\rho$ as:
\be
 {\cal H}' = {\cal U}\,{\cal H}\,{\cal U}^{\dagger}   ~~~~~~~~~
 \rho ' = {\cal U}\, \rho \,{\cal U}^{\dagger} ~, \label{eq:Hrhotr}
\ee
such that the partition function is invariant:
\be
{\rm e}^{-\beta F} = \tr {\rm e}^{-\beta {\cal H}} =
 \tr {\rm e}^{-\beta {\cal H}'}  ~.
\ee
Averages in the transformed system (based on $\rho '$) are denoted
as $\langle \cdots \rangle '$ and are related to the original
averages $\langle \cdots \rangle$ by
\be
\langle {\cal A} \rangle ' = \tr \rho ' {\cal A} =
\tr \,{\cal U}\, \rho \,{\cal U}^{\dagger} {\cal A} =
\tr \rho \,{\cal U}^{\dagger} {\cal A}\,{\cal U} =
\langle {\cal A}' \rangle ~,
\ee
with
\be
 {\cal A}' = {\cal U}^{\dagger} {\cal A}\,{\cal U} ~. \label{eq:optr}
\ee
${\cal U}$ locally rotates the spins around the $z$-axis
over an angle $\bq\cdot\br_j$. Using the properties of spin
operators, we have:
\be
 {s_{j}^{+}}' =  {\cal U}^{\dagger} s^{+}_j \,{\cal U} =
{\rm e}^{-i\bq\cdot\br_j} s^{+}_j ~,
\ee
such that (\ref{eq:s+tw}) changes into:
\be
\langle s^{+}_j \rangle ' = \langle {s^{+}_j} ' \rangle =
\ms {\rm e}^{i\bQ\cdot\br_j}  ~,
\ee
which puts the staggered magnetization in the new system
everywhere in the $x$-direction. Thus the constraint is simple
and homogeneous at the expense of changing the hamiltonian
from $\cal H$ to ${\cal H}'$.
It is however easy to compute the new Hubbard hamiltonian $\Hh '$ by
making use of the properties of the $c_{j \sigma}$ under the
transformation $\cal U$:
\be
{\cal U}\, c_{j \sigma}\, {\cal U}^\dagger =
{\rm e}^{-i\bq\cdot\br_j \sigma/2} c_{j \sigma} ~.
\ee
Using (\ref{eq:Hrhotr}), $\Hh '$ is:
\be
\Hh ' = - t \sum_{j,\delta,\sigma } {\rm e}^{i\sigma\bq\cdot\br_\delta/2}
c_{j+\delta,\sigma}^{\dagger} c_{j \sigma}
+ U \sum_{j}n_{j \uparrow}n_{j \downarrow} ,   \label{eq:Hhp}
\ee
where $\br_\delta = \br_{j+\delta} - \br_j$ is a vector connecting
nearest neighbors. Note that the new hamiltonian is again spatially
homogeneous and of the Hubbard form with a complex hopping integral
that depends on the pitch of the twist. So we must evaluate the free
energy of $\Hh'$ under the condition that the order parameter points
in the $x$-direction. In, e.g. the mean-field approximation one can
directly determine $F(\bq)$ and then expand $F(\bq)$ to obtain $\rs$
\cite{PDrhos}. Concrete calculations of $\rs$
starting from the
thermodynamic definition (using series expansions)
can be found in Refs.\cite{SH} and \cite{ShiSingh}
for the 2D Heisenberg antiferromagnet and Hubbard model, respectively.

In general we may evaluate $F(\bq)$ by expanding $\Hh'$ in powers
of $i\bq$:
\be
\Hh' = \Hh + i\bq\cdot\Hone + \ha i\bq\cdot\Htwo\cdot i\bq + \cdots
{}~, \label{eq:Hhp2}
\ee
with
\begin{eqnarray}
 \Hone = -\frac{t}{2} \sum_{j,\delta,\sigma} \, \br_\delta
 \sigma \, c_{j+\delta,\sigma}^{\dagger} c_{j \sigma}
 = N \, \Jhz \label{eq:H1} \\
 \Htwo = -\frac{t}{4} \sum_{j,\delta,\sigma} \, \br_\delta \br_\delta
    c_{j+\delta,\sigma}^{\dagger} c_{j \sigma} ~. \label{eq:H2}
\end{eqnarray}
We note that $\Hone$ is the current of the $z$-component of the spin and
that $\Htwo$ is (for cubic symmetry)
a diagonal tensor with the kinetic energy on the diagonal.
As $\bq$ is small, the free energy increment due to $\bq$ can be
calculated using perturbation theory using (\ref{eq:delF}):
\be
 F(\bq) - F({\bf 0}) = \ha i\bq \cdot
\left\{ \langle \Htwo \rangle - \beta \left( \Hone^\dagger ,
\Hone \right) \right\} \cdot i\bq
\ee
Here we use that $\langle \Hone \rangle = 0$ (no current in the unperturbed
state) and we note that $\Hone^\dagger = -\Hone$. So we obtain for $\rsbb$
the expression:
\be
\rsbb = -\frac{1}{N}
\left\{ \langle \Htwo \rangle^x +
\beta \left( \Hone , \Hone \right)^x \right\} ~.
\label{eq:rhosbb}
\ee
For cubic symmetry $\rsbb$ reduces to $\rsbb = \rs \Ibb$.
We stress, by using superscripts $x$,
that in (\ref{eq:rhosbb}) the averages are with respect to an
ensemble with the order parameter pointing in the $x$-direction
(and with $\Hh$ as hamiltonian).
\section{Equivalence of the two definitions for spin stiffness}
\renewcommand{\theequation}{5.\arabic{equation}}
\setcounter{equation}{0}
We now prove that the hydrodynamic definition (\ref{eq:divk2}) and the
thermodynamic definition (\ref{eq:Fq}) are equivalent. We do this
by considering the response to the perturbation:
\be
 \delta {\cal H} = - \sum_j \, (-1)^j \xi_j s^y_j ~, \label{eq:dHy}
\ee
where the $\xi_j$ are small and smoothly varying in space, e.g.
of the form:
\be
 \xi_j = \xi \, \cos \left( \bq\cdot\br_j \right) ~. \label{eq:xij}
\ee
The perturbation is added to the hamiltonian:
\be
 {\cal H} = \Hh - \xi^x N^x ~, \label{eq:Hhxin}
\ee
where we have introduced a small symmetry-breaking field $\xi^x N^x$ to
guarantee that the reference system has its order parameter pointing
in the $x$-direction.

Because of the intrinsic stiffness of the system the response to the
small perturbation (\ref{eq:dHy}) will be a local rotation of the
spins around the $z$-axis over a small angle $\fj$:
\be
 \langle s_j^x \rangle = (-1)^j \ms \cos \fj ~~~,~~~
 \langle s_j^y \rangle = (-1)^j \ms \sin \fj ~. \label{eq:sxsy2}
\ee
where $\fj$ is spatially smooth. The spatial variation (\ref{eq:xij})
will induce a similar wave pattern for the $\fj$:
\be
 \fj = A \, \cos \left( \bq\cdot\br_j \right) ~, \label{eq:phij}
\ee
with a small amplitude $A$ proportional to $\xi$.
Determining this proportionality is the main goal of this section,
as it leads directly to the singular behavior of the inner product
(\ref{eq:divk2}).
To see this we write $\delta {\cal H}$ with (\ref{eq:xij})
for the $\xi_j$ as:
\be
 \delta {\cal H} = -\ha \xi \left[ N^y(\bq) + N^y(-\bq) \right] ~.
\ee
The anticipated response (\ref{eq:sxsy2}) gives for the $y$-component of
$\bN(\bq)$:
\be
 \ha \langle N^y(\bq) + N^y(-\bq) \rangle =
\ms \sum_j \, \cos (\bq\cdot\br_j) \, \sin \fj ~.
\ee
Inserting (\ref{eq:phij}) and using that the angles $\fj$
are small yields:
\be
\langle N^y(\bq) + N^y(-\bq) \rangle =
2 \ms A \sum_j \, \cos^2 (\bq\cdot\br_j) = \Ms A  ~. \label{eq:MsA}
\ee
On the other hand, linear response theory (see (\ref{eq:delJ})) gives:
\be
\langle N^y(\bq) + N^y(-\bq) \rangle =
\frac{\beta\xi}{2} \left( N^y(\bq) + N^y(-\bq) , N^y(\bq) + N^y(-\bq) \right)
{}~. \label{eq:NyNy}
\ee
The modes $\bq$ and $-\bq$ have no overlap, so combining (\ref{eq:MsA})
and (\ref{eq:NyNy}) leads to:
\be
 \Ms A = \beta\xi \left( N^y(\bq) , N^y(\bq) \right) ~. \label{eq:MsA2}
\ee
Thus the computation of $A$ gives the inner product which must lead to the
hydrodynamic definition of $\rs$, (\ref{eq:divk2}).
We calculate the $\fj$ by first determining the increase in the
free energy for a set of arbitrary $\fj$ and then optimize this free
energy with respect to the $\fj$; this yields the $\fj$ resulting from
the $\xi_j$. As in the previous section, we gauge away the $\fj$ by an
unitary transformation:
\be
{\cal U} = {\rm e}^{i \sum_j \, \fj s^z_j}  ~. \label{eq:Ufj}
\ee
The averages in the new system are:
\begin{eqnarray}
\langle s^x_j \rangle ' & = &
\langle {\cal U}^{\dagger} s^x_j \,{\cal U} \rangle =
(-1)^j \ms  \label{eq:sxjt} \\
\langle s^y_j \rangle ' & = &
\langle {\cal U}^{\dagger} s^y_j \,{\cal U} \rangle = 0  ~, \label{eq:syjt}
\end{eqnarray}
such that in the new system the order parameter points in the
$x$-direction everywhere. This new constraint has to be combined with
the transformed hamiltonian:
\be
{\cal H}' = {\cal U} \left( \Hh - \xi^x N^x - \sum_j \, \xi_j (-1)^j
s^y_j \right) {\cal U}^\dagger ~.
\ee
We evaluate the free energy by perturbation theory. Therefore we
decompose ${\cal H}'$ into:
\be
{\cal H}' = \Hh + {\cal H}_{\rm a} + {\cal H}_{\rm b} ~, \label{eq:Htp}
\ee
where $\Hh$ is taken as the unperturbed hamiltonian together with the
constraints (\ref{eq:sxjt})-(\ref{eq:syjt}) on the order parameter.
As in the previous section, such averages are denoted by $\langle \cdots
\rangle^x$. The other two parts in (\ref{eq:Htp}) are given by:
\begin{eqnarray}
{\cal H}_{\rm a} & = & {\cal U} \, \Hh {\cal U}^\dagger - \Hh  \\
{\cal H}_{\rm b} & = & - {\cal U} \left[ \xi^x N^x + \sum_j \, \xi_j (-1)^j
s^y_j \right] {\cal U}^\dagger  ~.
\end{eqnarray}
These parts will be treated as perturbations on $\Hh$. ${\cal H}_{\rm a}$
can be expanded for small angles as:
\be
{\cal H}_{\rm a} = i \sum_j \, {\cal H}_j \fj - \ha \sum_{i,j} \,
\varphi_i {\cal H}_{ij} \fj + \cdots ~, \label{eq:Ha}
\ee
where it follows formally from (\ref{eq:Ufj}) that,
\begin{eqnarray}
{\cal H}_j & = & \left[ s^z_j , \Hh \right]  \\
{\cal H}_{ij} & = & \left[ s^z_i , \left[ s^z_j , \Hh \right] \right] ~.
\end{eqnarray}
For the Hubbard hamiltonian (cf. (\ref{eq:Hhp})) one can explicitly
transform $\Hh$:
\be
{\cal U} \, \Hh \, {\cal U}^\dagger =
- t \sum_{j,\delta,\sigma } {\rm e}^{i\sigma(\varphi_{j+\delta} - \fj)/2}
c_{j+\delta,\sigma}^{\dagger} c_{j \sigma}
+ U \sum_{j}n_{j \uparrow}n_{j \downarrow} ~,  \label{eq:Hhpfj}
\ee
which can then be expanded as in (\ref{eq:Ha}), yielding explicit expressions
for ${\cal H}_j$ and ${\cal H}_{ij}$.
The free energy associated with ${\cal H}_{\rm a}$ can be written as:
\be
 \delta F_{\rm a} = \ha \sum_{i,j} \, \varphi_i K_{ij} \fj ~,
\ee
with
\be
K_{ij} = - \langle {\cal H}_{ij} \rangle^x -
\beta \left( {\cal H}_i , {\cal H}_j \right)^x  ~.
\ee
This part of the free energy increase can be directly related to $\rsbb$ as
defined in the previous section. $K_{ij}$ is translationally invariant, i.e.,
a function of $\br_i - \br_j$. Moreover, we have:
\be
 K_{ij} = K_{ji}    ~~~~~~~~~~~,~~~~~~~~ \sum_j \, K_{ij} = 0 ~,
\label{eq:Kpr1}
\ee
as a constant rotation angle $\fj$ will not lead to an increase in
the free energy.

Angles with an uniform gradient,
\be
\fj = \bq\cdot\br_j ~,
\ee
have been discussed in the previous section and lead to $\rsbb$.
So:
\be
 \sum_{i,j} \,
\left( \bq\cdot\br_i \right) K_{ij} \left( \bq\cdot\br_j \right) =
 N\, \bq\cdot\rsbb\cdot\bq ~. \label{eq:Kpr2}
\ee
We will need the fourier transform of $K_{ij}$, which can be evaluated
with (\ref{eq:Kpr1}) and (\ref{eq:Kpr2}) as:
\be
\sum_{i,j} \, K_{ij} \, {\rm e}^{i\bq\cdot(\br_i - \br_j )} =
 N\, \bq\cdot\rsbb\cdot\bq ~~~~~~~\bq \rightarrow 0 ~. \label{eq:FtrK}
\ee

The part ${\cal H}_{\rm b}$ is found as:
\be
 {\cal H}_{\rm b} = - \sum_j \,
(-1)^j \left\{ \left[ \xi^x \cos \fj + \xi_j \sin \fj \right] s^x_j +
\left[ -\xi^x \sin \fj + \xi_j \cos \fj \right] s^y_j \right\} ~,
\ee
and yields:
\be
\delta F_{\rm b} = \langle {\cal H}_{\rm b} \rangle^x =
- \ms \sum_j \left( \xi^x \cos \fj + \xi_j \sin \fj \right) ~.
\ee
Thus the total increase in free energy is:
\be
\delta F = \delta F_{\rm a} + \delta F_{\rm b} =
\ha \sum_{i,j} \, \varphi_i K_{ij} \fj
- \ms \sum_j \left( \xi^x \cos \fj + \xi_j \sin \fj \right) ~.
\ee
Now the $\fj$ follow from the stationarity of $\delta F$,
which leads (for small $\fj$) to the linear equations:
\be
 \sum_i \, K_{ji} \varphi_i + \ms \xi^x \fj = \xi_j \, \ms ~.
\ee
Due to the translational invariance of $K_{ij}$ one finds that
indeed (\ref{eq:xij}) and (\ref{eq:phij}) form a solution with $A$ and
$\xi$ related by:
\be
 \left( K(\bq) + \ms \xi^x \right) \, A = \ms \xi ~,  \label{eq:Kq1}
\ee
where $K(\bq)$ follows from (\ref{eq:FtrK}):
\be
 K(\bq) = \bq\cdot\rsbb\cdot\bq ~~~~~~~\bq \rightarrow 0 ~. \label{eq:Kq2}
\ee
Now we can safely let $\xi^x \rightarrow 0$ and one sees that
(\ref{eq:MsA2}), (\ref{eq:Kq1}) and (\ref{eq:Kq2}) lead to:
\be
 \left( N^y(\bq), N^y(\bq) \right) =
\frac{\Ms^2}{\beta N\bq\cdot\rsbb\cdot\bq} ~~~~~~~\bq \rightarrow 0 ~,
\label{eq:hydrs}
\ee
which is a slight generalisation of (\ref{eq:divk2}) to systems
with a lower symmetry than the cubic symmetry.
Since indeed the same $\rsbb$ appears as in the previous section, we have
demonstrated the equivalence of the thermodynamic and hydrodynamic
definitions of the spin stiffness.

\section{Response to a Gauge Field}
\renewcommand{\theequation}{6.\arabic{equation}}
\setcounter{equation}{0}
A third aspect of $\rs$, which is unusual in the context of
antiferromagnetic order, is its role in the response to a
gauge field. This role is familiar in superconductivity, where the
presence of $\rs$ leads to special electromagnetic behavior in
a symmetry-broken state: the Meissner effect.
In a lattice system the electromagnetic field is introduced by
the so-called Peierls substitution \cite{Peierls}. It amounts to the
replacement,
\be
c_{j\sigma} \rightarrow c_{j\sigma} \exp \left[ -ie \int_0^{\br_j} \,
{\rm d}\br \cdot \bA (\br)/\hbar \right] ~,
\ee
where $e$ is the electric charge and $\bA(\br)$ the vector potential.
Executing this replacement in the Hubbard hamiltonian changes it into:
\be
{\cal H}_A =
- t \sum_{j,\delta,\sigma } c_{j+\delta,\sigma}^{\dagger} c_{j \sigma}
{\rm e}^{ie \bA_{j+\delta/2}\cdot\br_\delta/\hbar}
+ U \sum_{j}n_{j \uparrow}n_{j \downarrow} ~.  \label{eq:HhA}
\ee
We have introduced the abbreviation:
\be
\int_{\br_j}^{\br_{j+\delta}} \,{\rm d}\br \cdot \bA (\br) =
\bA_{j+\delta/2}\cdot\br_\delta  ~,
\ee
such that in the lattice version the vector potential $\bA_{j+\delta/2}$
is associated with the link between the sites $j$ and $j+\delta$.
By introducing the vector potential in this way one has to change the
expression for the electric current concomittantly. The current
$\bJ_{j+\delta/2}$ is also associated with the link $[j,j+\delta]$ and
it should obey the lattice version of charge conservation,
\be
\frac{\partial \rho_j}{\partial t} +
\sum_\delta \, \bJ_{j+\delta/2} \cdot \br_\delta = 0 ~,
\ee
with the charge density given by:
\be
\rho_j = e \, \sum_\sigma \, n_{j\sigma}  ~.
\ee
Using the equation of motion for $\rho_j$ one finds:
\be
\bJ_{j+\delta/2} = \frac{iet}{\hbar} \br_\delta \sum_\sigma
\left\{ c_{j+\delta,\sigma}^{\dagger} c_{j \sigma}
{\rm e}^{ie \bA_{j+\delta/2}\cdot\br_\delta/\hbar} -
c_{j\sigma}^{\dagger} c_{j+\delta , \sigma}
{\rm e}^{-ie \bA_{j+\delta/2}\cdot\br_\delta/\hbar} \right\}~.
\ee
When the system becomes superconducting the electric current $\bJ$
becomes proportional to $\bA$ (the London equation):
\be
 \bJ = - \left( \frac{n_{\rm s}e^2}{m} \right) \bA ~, \label{eq:LE}
\ee
where $n_{\rm s}$ is the density of the superconducting carriers
and $m$ is their mass. Superconductivity occurs in the negative-$U$
Hubbard model and it is the counterpart of the antiferromagnetism
in the positive-$U$ Hubbard model that we have been considering so far.
The two can be related to each other by the transformation
(see e.g. \cite{PDrhos}):
\be
c_{j\uparrow}' = c_{j\uparrow}  ~~~~~~,~~~~~~
c_{j\downarrow}' =(-1)^j \, c_{j\downarrow}^\dagger ~. \label{eq:sdphtr}
\ee
The transformation (\ref{eq:sdphtr}) transforms the charge density,
\be
\rho_j ' = e\, \sum_\sigma \, n_{j\sigma}' =
e\, \sum_\sigma \, \sigma n_{j\sigma} = 2e\, s^z_j ~,
\ee
into the $z$-component of the magnetization.
The hamiltonian ${\cal H}_A$ from (\ref{eq:HhA}) is transformed into:
\be
{\cal H}_A ' =
- t \sum_{j,\delta,\sigma } c_{j+\delta,\sigma}^{\dagger} c_{j \sigma}
{\rm e}^{ie \bA_{j+\delta/2}\cdot\br_\delta \sigma/\hbar}
+ U \sum_{j}n_{j \uparrow}n_{j \downarrow} ~.  \label{eq:HhAp}
\ee
Note the appearance of a factor $\sigma$ in the phase factor.
Comparing this expression with (\ref{eq:Hhpfj}), we see that they are
equivalent for:
\be
 \bA_{j+\delta/2}\cdot\br_\delta =
\left( \varphi_{j+\delta} -\fj \right) \frac{\hbar}{2e} ~.
\ee
The charge conservation law transforms into a conservation law for
the $z$-component of the magnetization:
\be
\frac{\partial s^z_j}{\partial t} +
\sum_\delta \, \bJ^z_{j+\delta/2} \cdot \br_\delta = 0 ~,
\ee
with the current $\bJ^z_{j+\delta/2}$ given by:
\be
\bJ^z_{j+\delta/2} = \frac{it}{2\hbar} \br_\delta \sum_\sigma
\left\{ c_{j+\delta,\sigma}^{\dagger} c_{j \sigma}
{\rm e}^{ie \bA_{j+\delta/2}\cdot\br_\delta \sigma/\hbar} -
c_{j\sigma}^{\dagger} c_{j+\delta , \sigma}
{\rm e}^{-ie \bA_{j+\delta/2}\cdot\br_\delta \sigma/\hbar} \right\} ~.
\ee
The equivalent of the London equation is the response of
$\bJ^z$ to $\bA$. We have two contributions: one because
$\bJ^z$ depends on $\bA$ and one because the hamiltonian
(\ref{eq:HhAp}) depends on $\bA$. Expansion of $\bJ^z$
gives:
\be
\bJ^z_{j+\delta/2} = \bJ^0_{j+\delta/2} +
 \frac{2e}{\hbar}\, \Jone_{j+\delta/2} \cdot \bA_{j+\delta/2}
  + \cdots ~,
\ee
with
\begin{eqnarray}
\bJ^0_{j+\delta/2} & = & \frac{it}{2\hbar} \br_\delta \sum_\sigma
\left( c_{j+\delta,\sigma}^{\dagger} c_{j \sigma} -
c_{j\sigma}^{\dagger} c_{j+\delta , \sigma} \right) \sigma \label{eq:J0} \\
\Jone_{j+\delta/2} & = & -\frac{t}{4\hbar}
 \br_\delta \br_\delta  \sum_\sigma
\left( c_{j+\delta,\sigma}^{\dagger} c_{j \sigma} +
c_{j\sigma}^{\dagger} c_{j+\delta , \sigma} \right) ~. \label{eq:J1}
\end{eqnarray}
Secondly, we have a contribution which results from the expansion
(\ref{eq:HhAp}) for small $\bA$:
\be
 \delta {\cal H}_A = -\frac{iet}{\hbar} \sum_{j,\delta,\sigma} \,
 c_{j+\delta,\sigma}^{\dagger} c_{j \sigma}
 \br_\delta \sigma \cdot \bA_{j+\delta/2} ~.
\ee
We note that $\Hone$ and $\Htwo$ as found in (\ref{eq:H1})-(\ref{eq:H2})
are related to $\bJ^0$ and $\Jone$
(Note that the two terms in the sum over $\sigma$ in (\ref{eq:J0}) and
(\ref{eq:J1}) contribute equally):
\begin{eqnarray}
 \Hone & = & -\frac{\hbar}{2i} \sum_{j,\delta} \, \bJ^0_{j+\delta/2}  \\
 \Htwo & = &  \frac{\hbar}{2}  \sum_{j,\delta} \,
              \Jone_{j+\delta/2}
\end{eqnarray}
The London equation (\ref{eq:LE}) only holds  for sufficiently
slowly varying $\bA_{j+\delta/2}$, i.e. the spatial variations
must be small over a coherence length. So we may take $\bA$ constant
and we then also have:
\be
 \delta {\cal H}_A = \frac{2ie}{\hbar} \, \Hone \cdot \bA ~.
\ee
The total magnetization current is defined as:
\be
 \bJ^z = \frac{1}{2N} \sum_{j,\delta} \, \bJ^z_{j+\delta/2} ~.
\ee
 From linear response theory we obtain for homogeneous $\bA$
(cf. (\ref{eq:delJ})):
\begin{eqnarray}
 \langle \bJ^z \rangle_{{\cal H} + \delta {\cal H}} & = &
 \frac{1}{2N} \left\{ \sum_{j,\delta} \, \frac{2e}{\hbar}
 \langle \Jone_{j+\delta/2} \rangle \cdot \bA - \beta
 \left(  \delta {\cal H}_A , \sum_{j,\delta} \, \bJ^0_{j+\delta/2} \right)
 \right\} \\
 ~~ & = & \frac{2e}{N\hbar^2} \left\{
\langle \Htwo \rangle + \beta \left( \Hone , \Hone \right) \right\}
\cdot \bA ~.
\label{eq:Jzav}
\end{eqnarray}
Comparing this with the expression (\ref{eq:rhosbb}) for $\rsbb$,
we may write:
\be
 \langle \bJ^z \rangle_{{\cal H} + \delta {\cal H}} =
 -\frac{2e}{\hbar^2} \, \rsbb \cdot \bA ~. \label{eq:Loneq}
\ee
Therefore indeed the same $\rsbb$ appears as before and we see that the
superconducting weight in the response to
a gauge field is also equivalent to the spin stiffness associated with
a twist in the order parameter. Very recently, a discussion of the
effect of a spin-dependent gauge field, as used in this section, was
given in Ref.\cite{KB} in the framework of Fermi liquid theory.

\section{Alternative forms for $\rs$}
\renewcommand{\theequation}{7.\arabic{equation}}
\setcounter{equation}{0}
In this section, we discuss some formal manipulations with the
expression for $\rs$. These are included because they involve some
subtleties due to the symmetry breaking.
The manipulations lead to expressions which are more general or easier
to interpret.
The formula (\ref{eq:rhosbb}) for $\rs$ together with the
definitions (\ref{eq:H1}) and (\ref{eq:H2}) refers explicitly to the
Hubbard model.
We can obtain a more general expression by expanding $\cal U$,
as given by (\ref{eq:U}) in powers of $i\bq$:
\be
 {\cal U} = {\rm e}^{i\bq\cdot\thz} = 1 + i\bq\cdot\thz + \cdots ~,
\label{eq:Uthz}
\ee
with $\thz$ given by:
\be
 \thz = \sum_j \, \br_j s_j^z ~.
\ee
Inserting (\ref{eq:Uthz}) in the definition of ${\cal H}'$ yields
a similar expansion as (\ref{eq:Hhp2}) with:
\be
 \Hone = \left[ \thz, \Hh \right] ~~~~~~~~~~
 \Htwo \, = \left[ \thz , \left[ \thz , \Hh \right] \right] ~.
\label{eq:H1H27}
\ee
Using this in the expression (\ref{eq:rhosbb}) gives for $\rsbb$:
\be
 \rsbb = -\frac{1}{N} \left\{
\langle \left[ \thz , \left[ \thz , \Hh \right] \right] \rangle^x +
\beta \left( \left[ \thz , \Hh \right] , \left[ \thz , \Hh \right]
\right)^x \right\} ~. \label{eq:rhosthz}
\ee
This expression makes no reference to the Hubbard model and is therefore
generally valid for lattice hamiltonians.
We have augmented the average and the
inner product with a superscript $x$ to reflect that the averages are
taken in a system where the order parameter points in the $x$-direction.
Such a warning signal is not superfluous as the following ``derivation''
may show. Suppose we use the relation,
\be
 \dthz = \frac{i}{\hbar} \left[ {\cal H} , \thz \right] ~, \label{eq:thd}
\ee
to write for the second term in (\ref{eq:rhosthz}):
\be
 \left( \left[ \thz, {\cal H} \right] ,
\left[ \thz, {\cal H} \right] \right) =
 -\frac{\hbar}{i} \left( \left[ \thz, {\cal H} \right] , \dthz \right) ~,
\label{eq:inthz1}
\ee
and use the relation (\ref{eq:timederid}) to transform it into:
\be
 \left( \left[ \thz, {\cal H} \right] ,
\left[ \thz, {\cal H} \right] \right) =
 -\frac{1}{\beta} \langle \left[ \thz,
\left[ \thz , {\cal H} \right] \right]
\rangle ~. \label{eq:inthz2}
\ee
Then we would discover that the two contributions in (\ref{eq:rhosthz})
exactly compensate each other! The error in this reasoning is that
we have ignored the fact that the averages have to be taken in a
symmetry-broken state. However, relation (\ref{eq:timederid}) is based
on a cyclic rotation of the operators in a trace, which is only
permitted when the trace is taken over the whole Hilbert space.

We can elucidate this point further by implementing the constraint
on the order parameter by including a symmetry-breaking term in the
hamiltonian as in (\ref{eq:Hhxin}). Then we would have for
(\ref{eq:thd}):
\be
\left[ \thz , \Hh \right] = \left[ \thz , {\cal H}  \right]
 + \xi^x \, \left[ \thz , N^x \right] =
- \frac{\hbar}{i} \dthz + i \xi^x \, \sum_j \, (-1)^j \br_j s^y_j ~.
\label{eq:thzHh}
\ee
The extra term looks innocent, because $\xi^x$ is vanishingly small,
but it is not, as we shall see. Using (\ref{eq:thzHh}), gives for
the second term in (\ref{eq:rhosthz}):
\be
 \left( \left[ \thz, \Hh \right] ,
\left[ \thz, \Hh \right] \right)^x =
 -\frac{1}{\beta} \langle \left[ \thz,
\left[ \thz , \Hh \right] \right] \rangle^x + i \xi^x
\left( \left[ \thz , \Hh \right] ,
\sum_j \, (-1)^j \br_j s^y_j \right)^x ~. \label{eq:mani}
\ee
Using (\ref{eq:thzHh}) again in the first entry of the second term
of (\ref{eq:mani}),
we obtain two contributions, one proportional to $\xi^x$ and one to
$\left( \xi^x \right)^2$. The former reads with the use of
(\ref{eq:timederid}):
\be
i \xi^x \frac{i}{\hbar}
\left( \dthz , \sum_j \, (-1)^j \br_j s^y_j \right)^x =
\frac{\xi^x}{\beta\hbar^2} \left( \sum_j \, \br_j \br_j \right) \ms ~.
\ee
The lattice sum can take arbitrarily large values and thus one cannot
rely on the smallness of $\xi^x$ to ignore its contribution. The
term proportional to $\left( \xi^x \right)^2$ is equally difficult
to interpret.

We may however use these types of operation to give the expression for
$\rs$ yet another form, and thereby bring this subtle point into
focus. First, we relate the second part of (\ref{eq:rhosthz}) to
a current-current inner product:
\be
\left( \left[ \thz, \Hh \right] ,
\left[ \thz, \Hh \right] \right)^x = \hbar^2 N^2
\left( \Jhz , \Jhz \right)^x
{}~,
\ee
with $\Jhz$ defined in (\ref{eq:H1}). Next, we use the identity,
\be
 \left[ \thz, \left[ \thz , \Hh \right] \right] =
 \left[ \thx, \left[ \thx , \Hh \right] \right]  ~, \label{eq:thzthx}
\ee
which is a direct consequence of the magnetic isotropy and which
is easily proven for the Hubbard model by direct evaluation.
Now for the right-hand side of (\ref{eq:thzthx}) we can apply the train
of arguments (\ref{eq:thd})-(\ref{eq:inthz2}) everywhere replacing
$z$ by $x$, since $\thx$ commutes with the symmetry-breaking term.
So (\ref{eq:inthz2}) {\em holds} for the $x$-component.
This permits us to write the expression for $\rs$ in the form:
\be
 \rsbb = \beta\hbar^2 N \left\{
\left( \bJ^\parallel_h , \bJ^\parallel_h \right) -
\left( \bJ^\perp_h , \bJ^\perp_h \right) \right\} ~,
\label{eq:parperp}
\ee
where we have used the coordinate-free notation: parallel and perpendicular
are to be understood with respect
to the orientation of the order parameter. This expression most
clearly shows that $\rs$ is induced by symmetry breaking: without
symmetry breaking the distinction between parallel and perpendicular
disappears and $\rs$ vanishes. In a mean-field (or: BCS) approximation
one finds at $T=0$ only the parallel contribution. For increasing
$T$ a gradual compensation occurs between the two terms, which becomes
complete at $T=T_{\rm c}$.

Since (\ref{eq:thzthx}) relies on the magnetic isotropy of the model,
the expression (\ref{eq:parperp}) is no longer valid when the isotropy
is broken by an external magnetic field in the $z$-direction. Then
(\ref{eq:rhosthz}) still holds and also the Halperin-Hohenberg relation
(\ref{eq:HHrel}) remains valid. Of course both $\rs$ and $\chi_\perp$ are
affected by the presence of such a magnetic field, as can e.g. be seen
from a mean-field treatment of these quantities \cite{PDrhos,PDvsw}.

As a final comment on this genre of expressions we discuss the Bogoliubov
inequality which is sometimes used to make (\ref{eq:divk2}) plausible
\cite{Fors}. The general form reads:
\be
\left( {\cal A} , {\cal A} \right) \left( {\cal B} , {\cal B} \right)
\geq \mid \left( {\cal A} , {\cal B} \right) \mid^2 ~. \label{eq:Bineq}
\ee
Take ${\cal A} = N^y (\bk)$ and ${\cal B} = \dot{M}^z(\bk)$ such that
$\left( {\cal A} , {\cal A} \right)$ is the desired inner product. The
right-hand side of (\ref{eq:Bineq}) is then given by (\ref{eq:NyMzd}).
One might think that $\left( {\cal B} , {\cal B} \right)$ is associated
with the inner product $\left( \bJ^z ,\bJ^z \right)$,
since for small $\bk$ one has:
\be
 \dot{M}^z(\bk) \simeq \frac{\bk}{\hbar} \cdot \left[ \thz, {\cal H} \right]
 = i N \bk \cdot \bJ^z ~.  \label{eq:Mdot}
\ee
This is not correct as a more precise analysis shows. First we use
(\ref{eq:timederid}) to write:
\be
 \left( \dot{M}^z(\bk) , \dot{M}^z(\bk) \right) = \frac{i}{\beta\hbar}
\langle \left[ \dot{M}^z(-\bk) , M^z(\bk) \right] \rangle ~. \label{eq:MdMd}
\ee
Then we expand the first entry as in (\ref{eq:Mdot}), but taking the
symmetry-breaking term explicitly into account:
\be
 \dot{M}^z(-\bk) \simeq
-\frac{\bk}{\hbar} \cdot \left[ \thz, \Hh \right] -
\frac{\xi^x}{\hbar} \left( N^y(-\bk) - N^y \right) ~.
\ee
The small symmetry-breaking term gives:
\be
-\frac{i\xi^x}{\beta\hbar^2} \langle \left[ N^y(-\bk) - N^y , M^z(\bk)
\right] \rangle = \frac{\xi^x \Ms}{\beta\hbar^2}
\ee
Then we obtain for small $\bk$ for (\ref{eq:MdMd}):
\be
\left( \dot{M}^z(\bk) , \dot{M}^z(\bk) \right) = \frac{1}{\beta\hbar^2}
\left\{ -\bk \cdot
\langle \left[ \thz, \left[ \thz , \Hh \right] \right] \rangle
\cdot \bk + \xi^x \Ms \right\} ~.
\ee
Substituting all of this in (\ref{eq:Bineq}) we have:
\be
 \left( N^y(\bk), N^y(\bk) \right) \geq
\frac{\Ms^2}{\beta \left(  -\bk \cdot
\langle \left[ \thz, \left[ \thz , \Hh \right] \right] \rangle
\cdot \bk + \xi^x \Ms \right)} ~. \label{eq:Bineq2}
\ee
This result should be compared to (\ref{eq:hydrs}); substituting for
$\rsbb$ the expression (\ref{eq:rhosbb}), with (\ref{eq:H1H27})
for $\Hone$ and $\Htwo$, one sees that the Bogoliubov inequality above
only involves the double commutator, or, equivalently, the
inner product of parallel currents, and not the inner product of
perpendicular currents.
Note that (\ref{eq:Mdot})
seems to suggest that the inner product of perpendicular currents
would be involved. This is not the case and generally we are indeed
faced with an {\em inequality} (the symmetry-breaking term
is innocent in this respect).

\section{Discussion}
\renewcommand{\theequation}{8.\arabic{equation}}
\setcounter{equation}{0}
In the foregoing sections we have discussed the role of $\rs$:
\begin{itemize}
\item as a parameter in the hydrodynamic (spin wave) equations,
\item as the parameter determining the increase in the free
energy due to long-wavelength variations in the phase of
the order parameter,
\item as the proportionality parameter between a perturbing gauge
field and the induced current.
\end{itemize}
By explicit calculation we have shown that it is the same $\rs$
entering in all these aspects. We have placed the discussion in
the context of the Hubbard model which allows more explicit
expressions. As we have indicated, the expressions can easily
be extended to more general lattice models.

In the case of magnetic isotropy we can write $\rs$ as the
difference of two current-current inner products containing the
spin currents parallel and perpendicular to the order parameter.
The Bogoliubov inequality only involves the parallel current.
In the BCS approximation it becomes an equality only at $T=0$.

A final comment we want to make concerns the issue of gauge invariance.
The London equation (\ref{eq:LE})
can be extended to the frequency-wavevector
domain as:
\be
 \langle \bJ (\bk,\omega) \rangle =
 \chbb (\bk,\omega) \cdot \bA (\bk,\omega) ~,
\ee
where $\bA (\bk,\omega)$ is the double fourier transform of
$\bA (\br,t)$ with respect to space and time and $\chbb (\bk,\omega)$
a generalized susceptibility (which has the role $\rs$ had before).
The limit of slow variations in
space and time is delicate \cite{SWZ}.
In the limit of first $\bk \rightarrow 0$
and then $\omega \rightarrow 0$ one has electric response and
$\chbb$ leads to the Drude weight. In the opposite order, first
$\omega \rightarrow 0$ and then $\bk \rightarrow 0$, one has magnetic
response leading to a non-vanishing superconducting weight, which
is considered here. Electromagnetic gauge invariance requires that a purely
longitudinal $\bA (\bk,\omega) = \bk a(\bk,\omega)$ has no response,
since it can be gauged away.
Thus:
\be
 \chbb (\bk,\omega) \cdot \bk = 0 ~. \label{eq:Couga}
\ee
We have been considering static response throughout, so we have
taken the limit $\omega \rightarrow 0$ first and we have not been
worrying about the $\bk$-dependence of the spin stiffness $\rsbb$.
Thus the expressions given
for $\rsbb$ do not obey the condition (\ref{eq:Couga}). The origin
of this paradox is not to be traced to shortcomings of e.g. the
BCS-approximation, as is sometimes done in the idea that vertex
corrections will restore gauge invariance \cite{SWZ}. The answer comes
from the change in gauge invariance in the symmetry-broken state.
In section 6, we saw that gauge fields are similar to imposed twists
on the phase of the order parameter. The theory is required to
be invariant under the combined gauge transformation:
\begin{eqnarray}
 \bA_{j+\delta/2}' \cdot \br_\delta & = &
\bA_{j+\delta/2} \cdot \br_\delta + \chi_{j+\delta} - \chi_j  \\
 \phi_j ' & = &  \phi_j + \frac{2e}{\hbar} \chi_j ~,
\end{eqnarray}
where $\phi_j$ is the phase of the order parameter and $\chi_j$
an arbitrary function. Correspondingly the London equation
(\ref{eq:Loneq}) should be formulated in a gauge-invariant way as:
\be
 \langle \bJ \rangle_{{\cal H} + \delta {\cal H}} =
 -\frac{2e}{\hbar^2} \, \rsbb \cdot
 \left(\bA - \frac{\hbar}{2e} {\bf \nabla} \phi \right)~.
\ee
Then the requirement of gauge invariance is automatically fulfilled
and it imposes no further requirements on $\rsbb$.


\newpage
\appendix
\section*{Appendix}
\renewcommand{\theequation}{A.\arabic{equation}}
\setcounter{equation}{0}
In this appendix we derive expressions for the change in the average
of an operator $\cal J$ and the change in free energy upon perturbing the
hamiltonian ${\cal H}$ by adding an operator $\delta {\cal H}$.
These expressions, up to first and second order in $\delta {\cal H}$,
respectively, are given in equations (\ref{eq:delJ}) and (\ref{eq:delF})
in terms of the inner product defined by (\ref{eq:inp}) and are
repeated here:
\begin{eqnarray}
 \delta \langle {\cal J} \rangle & \equiv &
\langle {\cal J} \rangle_{{\cal H} + \delta {\cal H}} -
\langle {\cal J} \rangle =
- \beta \left( \delta {\cal H}^{\dagger}, {\cal J} \right) ,
\label{eq:delJA} \\
 \delta F & \equiv & -\frac{1}{\beta} \left[ \ln \tr{\rm e}^{-\beta (
{\cal H} + \delta {\cal H})} -  \ln \tr{\rm e}^{-\beta {\cal H}} \right]
= \langle \delta {\cal H} \rangle - \frac{\beta}{2}
\left(\delta {\cal H}^{\dagger}, \delta {\cal H} \right) .
\label{eq:delFA}
\end{eqnarray}
Unless explicitly indicated averages are taken with respect to the
operator ${\rm e}^{-\beta {\cal H}}$ (see (\ref{eq:ave})).

The derivation of both equations proceeds by way of the operator
identity:
\be
{\rm e}^{-\beta ({\cal H} + \delta {\cal H})} =
{\rm e}^{-\beta {\cal H}} - \int_0^{\beta} {\rm d} \lambda \,
{\rm e}^{-(\beta - \lambda) {\cal H}} \delta {\cal H} \,
{\rm e}^{- \lambda ({\cal H} + \delta {\cal H})} , \label{eq:opid}
\ee
which is proven by noting that both left- and right-hand side are
equal for $\beta = 0$ and have identical derivatives with respect to
$\beta$.

Equation (\ref{eq:delJA}) is now derived by inserting (\ref{eq:opid})
into:
\be
\langle {\cal J} \rangle_{{\cal H} + \delta {\cal H}} =
\frac{\tr{\rm e}^{- \beta ({\cal H} + \delta {\cal H})} {\cal J}}
     {\tr{\rm e}^{- \beta ({\cal H} + \delta {\cal H})}} ,
\ee
expanding to first order in $\delta {\cal H}$ and using the following
result (which is easily proved by cyclic permutation in the trace):
\be
\int_0^{\beta} {\rm d} \lambda \langle \delta {\cal H}(\lambda) \rangle
\equiv \int_0^{\beta} {\rm d} \lambda \langle {\rm e}^{\lambda {\cal H}}
\delta {\cal H} \,{\rm e}^{-\lambda {\cal H}} \rangle =
\beta \langle \delta {\cal H} \rangle ~. \label{eq:intdelH}
\ee

To derive (\ref{eq:delFA}), we start by iterating (\ref{eq:opid}) once
and expanding to second order in $\delta {\cal H}$:
\be
\tr{\rm e}^{- \beta ({\cal H} + \delta {\cal H})} =
\tr{\rm e}^{- \beta {\cal H}} \left( 1 -
\int_0^{\beta} {\rm d} \lambda \langle \delta {\cal H}(\lambda) \rangle +
\int_0^{\beta} {\rm d} \lambda \int_0^{\lambda} {\rm d} \lambda'
\langle \delta {\cal H}(\lambda) \delta {\cal H}(\lambda') \rangle +
\cdots \right) ~.  \label{eq:expdH}
\ee
Before taking the logarithm in order to compute $\delta F$ it is useful
to prove the identity:
\be
\int_0^{\beta} {\rm d} \lambda \int_0^{\lambda} {\rm d} \lambda'
\langle \delta {\cal H}(\lambda) \delta {\cal H}(\lambda') \rangle =
\frac{\beta}{2} \int_0^{\beta} {\rm d} \lambda
\langle \delta {\cal H}(\lambda) \delta{\cal H} \rangle ~. \label{eq:ddH}
\ee
The proof proceeds by a sequence of changes of variable in the
integral, introducing the notation:
\be
 f(\tau) = \langle \delta {\cal H}(\tau) \delta{\cal H} \rangle ~,
\ee
and using:
\be
 f(\beta - \tau) = f(\tau) ~. \label{eq:fbtau}
\ee
Equation (\ref{eq:fbtau}) is again proved by cyclical permutations in the
trace. Explicitly:
\begin{eqnarray}
& ~ & \int_0^{\beta} {\rm d} \lambda \int_0^{\lambda} {\rm d} \lambda'
\langle \delta {\cal H}(\lambda) \delta {\cal H}(\lambda') \rangle =
\int_0^{\beta}  {\rm d} \lambda \int_0^{\lambda} {\rm d} \lambda'
\langle \delta {\cal H}(\lambda-\lambda') \delta {\cal H} \rangle =
\nonumber \\
& ~ & \int_0^{\beta} {\rm d} \lambda \int_0^{\lambda} {\rm d}
\tau \, f(\tau) =
\int_0^{\beta} {\rm d} \tau \int_{\tau}^{\beta} {\rm d}
\lambda \, f(\tau) =
\int_0^{\beta} {\rm d} \tau \,(\beta - \tau) f(\tau) =  \nonumber \\
& ~ & \int_0^{\beta} {\rm d} \tau \, \tau \, f(\beta -\tau) =
 \frac{1}{2} \int_0^{\beta} {\rm d} \tau \,(\beta - \tau + \tau) f(\tau) =
\frac{\beta}{2} \int_0^{\beta} {\rm d} \tau \,
\langle \delta {\cal H}(\tau) \delta{\cal H} \rangle ~.~~ \nonumber
\end{eqnarray}
Now equation (\ref{eq:delFA}) is easily derived by inserting (\ref{eq:ddH})
and (\ref{eq:intdelH}) into (\ref{eq:expdH}), taking the logarithm,
expanding to second order in $\delta {\cal H}$, and recognizing in the
final result the inner product (\ref{eq:inp}).


\newpage
\begin{thebibliography}{99}
\bibitem{Fors} Forster D 1975 {\em Hydrodynamic Fluctuations,
Broken Symmetry, and Correlation Functions} (W.A. Benjamin)
\bibitem{SWZ} Scalapino D J, White S R and Zhang S C 1992
Phys. Rev. Lett. {\bf 68} 2830; 1993
Phys. Rev. B {\bf 47} 7995
\bibitem{HalHo} Halperin B I and Hohenberg P C 1969 Phys. Rev. {\bf 188} 898
\bibitem{FBJ} Fisher M E, Barber M N and Jasnow D 1973 Phys. Rev. A {\bf 8}
1111
\bibitem{Hub} Hubbard J 1963 Proc. Roy. Soc. (London) A{\bf 276} 238
\bibitem{Kubo} Kubo R 1957 J. Phys. Soc. Japan {\bf 12} 570
\bibitem{Kruger} Kr\"{u}ger P and Schuck P 1994 Europhys. Lett. {\bf 27} 395
\bibitem{PDrhos} Denteneer P J H, An Guozhong and van Leeuwen J M J 1993
Phys. Rev. B {\bf 47} 6256
\bibitem{SH} Singh R R P and Huse D 1989 Phys. Rev. B {\bf 40} 7247
\bibitem{ShiSingh} Shi Z-P and Singh R R P 1995 Europhys. Lett. {\bf 31} 219
\bibitem{Peierls} Peierls R 1933 Z. Phys. {\bf 80} 763
\bibitem{KB} Bedell K S and Farinas P F 1995 Phys. Rev. Lett. {\bf 74} 4285
\bibitem{PDvsw} Denteneer P J H and van Leeuwen J M J 1993 Europhys. Lett.
{\bf 22} 413
\end{thebibliography}
\end{document}